    \documentclass[12pt,psf,epsf]{article}
\usepackage[centertags]{amsmath}
\usepackage{amssymb}
\usepackage{graphicx}
\usepackage{epsfig}
\usepackage{ulem}
\usepackage[english]{babel}
\usepackage{array}
\usepackage{amsthm}
\usepackage{latexsym}
\usepackage[mathcal]{euscript}
\pdfoutput=1
\usepackage{epsfig}

 \usepackage{jheppub}
 \usepackage{hyperref}

\usepackage{subfigure}
\usepackage{psfrag}

\usepackage{epsfig}
\usepackage[latin1]{inputenc}
\usepackage{float}
\usepackage{graphicx}
\usepackage{cancel}
\usepackage{mathrsfs}
\usepackage{amssymb}
\usepackage{amsfonts}
\usepackage{amsmath}
\usepackage{slashed}
\usepackage{graphicx}
\usepackage{bm}
\usepackage{color}
\newcommand{\be}{\begin{equation}}
\newcommand{\ee}{\end{equation}}
\newcommand{\bea}{\begin{eqnarray}}
\newcommand{\eea}{\end{eqnarray}}
\newcommand{\bwt}{\begin{widetext}}
\newcommand{\ewt}{\end{widetext}}

\newcommand{\nn}{\nonumber}

\newcommand{\bi}{\begin{itemize}}
\newcommand{\ei}{\end{itemize}}

\usepackage{setspace}

\begin{document}

\title {When things stop falling, chaos is suppressed}

\author{Dmitry S. Ageev, Irina Ya. Aref'eva}
\affiliation{Steklov Mathematical Institute, Russian Academy of Sciences, Gubkin str. 8, 119991
Moscow, Russia}
\emailAdd{ageev@mi-ras.ru}
\emailAdd{arefeva@mi-ras.ru}

\abstract{
 This note is devoted to the investigation of Susskind's proposal \cite{Susskind1} concerning the correspondence between the operator growth in chaotic  theories and the radial momenta of the particle  falling in the AdS black hole. We study this proposal and consider the simple example of an operator with the global charge described by the charged particle falling to the Reissner-Nordstrom-AdS black hole. Different charges of the particle lead  to qualitatively different behavior  of the particle momenta and consequently  change of the operator size behavior. This holographic result is supported by  different examples of chaotic models with a finite chemical potential where the suppression of chaos has been observed.

}

\maketitle

\newpage

\section{Introduction}
In the recent years the quantum chaotic systems and their connection to the gravity has attained a lot of attention. The systematic investigation of this correspondence revealed new concepts like scrambling, quantum complexity  \cite{Susskind2}-\cite{Size1}. The notion of operator size, roughly speaking, describes how many operators taken as  "basis operators" are involved in the description of evolution after perturbation. It was shown, that in the chaotic systems this growth is exponential and its power is called the Lyapunov exponent. In the AdS/CFT correspondence, the operators in the quantum theory can be related to the particle in the bulk and in the context of "GR=QM" proposal 
it was found that this exponential operator size growth on the gravity side corresponds to the (exponential) growth of the radial momentum of the particle falling under the black hole horizon \cite{Susskind1,Susskind2}.

In this paper we investigate the above mentioned  GR=QM proposal \cite{Susskind1,Susskind2}  and the correspondence between the operator growth in chaotic holographic theories and the radial particle momenta by a simple example of an operator with a global charge described by a charged particle falling in the Reissner-Nordstrom-AdS black hole. We find that the different  charges of the particle, lead  to qualitatively different behaviour of the particle. We interpret this difference as a supression of a chaotic behaviour in a systems with finite chemical potential. Also we discuss the quantum mechanical model exhibiting the similar behaviour.

The paper is organized as follows. First we introduce the main ingredients of the correspondence. In the third section we compute the evolution of momentum growth for charged particles numerically solving equations of motion. In the fourth section we discuss the SYK model at finite chemical potential and future directions of investigation. We end with the discussion.

\section{Charge falling in the Reissner-Nordstrom-AdS black hole}  
Let us consider the planar $d$-dimensional Reissner-Nordstrom-AdS black hole with the outer horizon at $z_h$. This geometry is  given by the metric 
\bea \label{metr-1}
&&ds^2=\frac{1}{z^2}\left(-f(z)dt^2+\frac{dz^2}{f(z)}+d\bar x^2 \right)\\
&&f(z)=1-M \Big(\frac{z}{z_h}\Big)^d+Q\Big(\frac{z}{z_h}\Big)^{2d-2},
\eea
where the gauge field supporting this solution is
\be\label{metr-2}
A_t=\mu\Big(1-\Big(\frac{z}{z_h}\Big)^{d-2}\Big),
\ee
and parameters $M$, $Q$ are defined to be
\bea\label{metr-3}
&&M=1+z_h^2\mu^2,\\
&&Q=z_h^2\mu^2,\\
&&T=\frac{d-(d-2) \mu ^2 z_h^2}{4 \pi  z_h},\\
&&z_h=\frac{d + 2 Q^2 - d \,Q^2}{4 \pi T}.
\eea
Here $T$ is the temperature of the black hole and $\mu$ is the chemical potential in the dual quantum field theory living on the boundary $z=0$. 
  Following \cite{Susskind2} there are different regions  in the charged near-extremal black hole. The first one is the  near-horizon or Rindler region that corresponds to the universal late time chaos onset at $t \rightarrow \infty $. The other part of the space $z>0$ consists of the throat region and the Newtonian one is the closest to the boundary. In this paper we concentrate our attention on the Rindler region. 

The proposal of Susskind \cite{Susskind1} states the correspondence between radial component of momentum $p_z$ of the neutral particle falling under the black hole horizon and the growth of the operator size. This correspondence is formulated as
\be 
\text{\bf Operator size} \longleftrightarrow {\bf p_z(t)},\label{I}
\ee
 where the operator size can be defined as follows. If we define the perturbation of a boundary theory by the time-evolving operator $W(t)$ and expand it in the basis of some elementary operators $\psi_{a_s}$
\be 
W(t)=\sum_{s}W_s(t),\,\,\,\,\,\,W_s=\sum_{a_1<...<a_s} c_{a_1...a_s}(t)\psi_{a_1}...\psi_{a_s}
\ee
then $s$  is called the operator size (see for example \cite{Size1,Shenker:2014cwa}). In \cite{Susskind1,Susskind2} it was shown, that  in a theory dual to the $d$-dimensional black hole, the neutral operator $W$   corresponds to the massive particle in the bulk. At a late times this momenta (i.e. when the particle is in the near  horizon zone) exhibits exponential growth
\be 
p_z(t) \sim e^{\frac{2\pi}{\beta}t}
\ee
and consequently the operator size also growth exponentially, where  $\beta$ is the inverse temperature.

In this framework it is natural to assume, that charged particle falling in the charged black hole background corresponds to evolution of the charged operator after a perturbation of the system at finite chemical potential.
The action of the charged particle has the  form
\bea \label{Sq}
S=-m\int\sqrt{-g_{\mu\nu}\dot x^{\mu} \dot x^{\nu}}d\tau+q A_{\mu}\dot x^{\mu}d\tau,
\eea
where the momentum is defined as
\bea
&&p_{\mu}=m \frac{g_{\mu\nu}\dot x^{\nu} }{\sqrt{-g_{\mu\nu}\dot x^{\mu}\dot x^{\nu}}}+q A_{\mu}.
\eea 
In the following we take $m=1$ without loss of generality.
If we take charge $q$ of the same sign as $Q$ (i.e. positive) then after some critical value $q_{crit}$ the qualitative behaviour of the particle changes. If $q<q_{crit}$ particle falls under the horizon, while for $q>q_{crit}$ it oscillates between the boundary and horizon.

\section{Charge stops the fall}
Let us take a closer look  on the motion of the particle with a charge  \eqref{Sq} in the RN-AdS black hole given by \eqref{metr-1}.

\begin{figure}[h!]
\centering
\includegraphics[width=8cm]{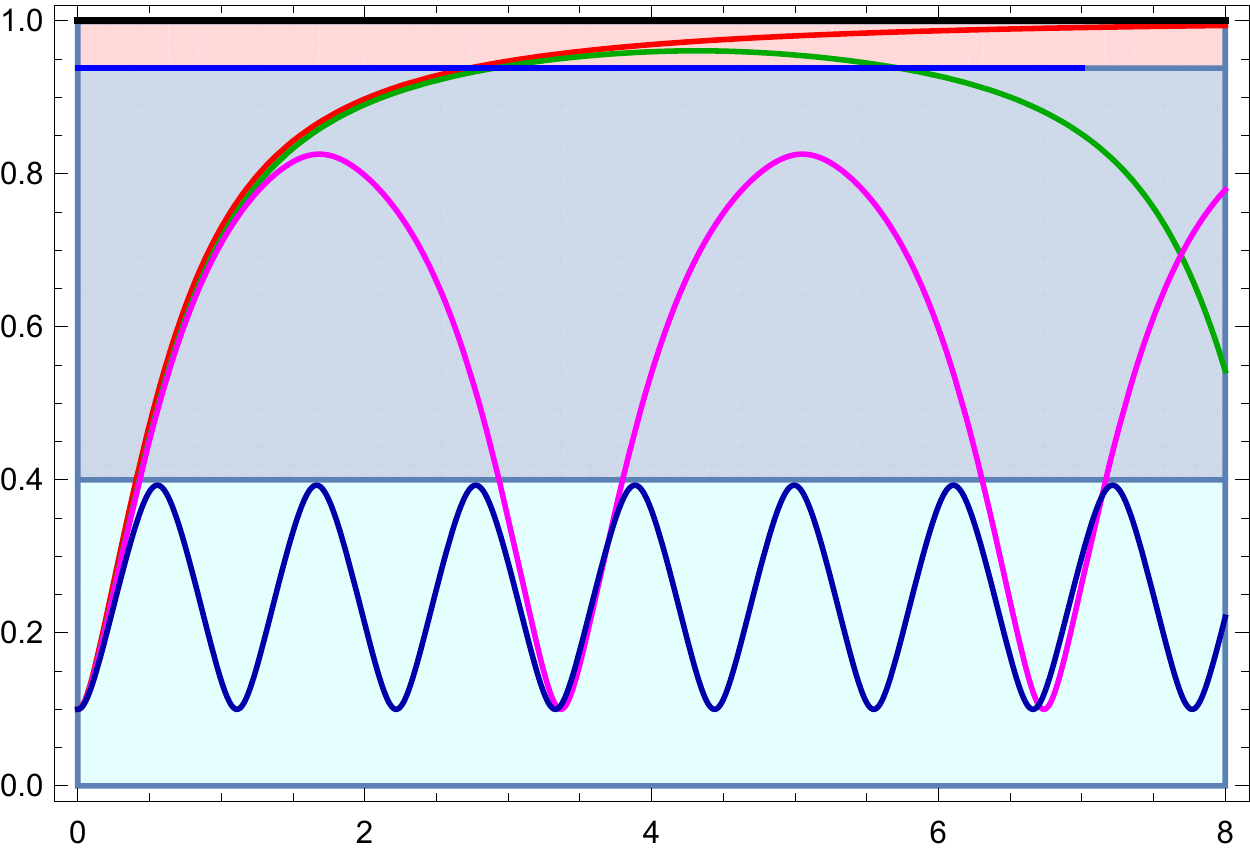}
 \caption{Three regions of the black hole for  the fixed chemical potential $\mu=1.635<\mu_{extr}$ (here $\mu_{extr}=1.732$ is the value of the chemical potential  corresponding to the extremal black hole). The pink region near the horizon $z_h=1$  is the Rindler region, the middle blue region corresponds to the throat and the cyan one is the outer region.  
   The red curve corresponds to $q=0$, the green curve corresponds to $q<q_{R/T}$ ($q_{R/T}$ is the value of the charge at which the particle starting from the given point, say $z_*=0.1$, touches
 the boundary of the Rindler and the throat regions),  the blue one to $q=q_{T/O}$, the value of the charge at which the particle is restricted to the outer region. The magenta curve corresponds to  the particle worldline crossing the blue and the cyan regions and its charge $q$
 satisfies $q_{R/T}<q<q_{T/O}$.}
 \label{fig:world}
\end{figure}

Taking the particle worldline parametrization $z=z(t)$ we get the action
\be
S=-\int\frac{\sqrt{f(z)-\dot z^2/f(z)}}{z}dt+ q\mu\Big(1-\frac{z^{d-2}}{z_h^{d-2}}\Big)dt.
\ee
The energy $E$ is
\bea
&&E=-\mu q\Big(1-    \frac{z^{d-2}}{z_h^{d-2}}\Big)+\frac{f(z)^{3/2}}{z(t)
   \sqrt{f(z)^2-\dot z^2}}.
\eea
 We consider the case of the positive particle charge  $q>0$.   When the energy is negative, the particle oscillates between two turning points $z_{*,\pm}$. In this case  the charge of the particle is larger than the critical one given by 

\be
q_{crit}=\frac{\sqrt{f(z_{*,-})}}{ z_* A_t(z_{*,-})},
\ee
where $A_t$ is given by \eqref{metr-2}.

We  take $d=3$ for definiteness and solve the equations of motion corresponding to \eqref{Sq} numerically and plot the particle worldline $z(t)$ in Fig.\ref{fig:world} for different values of the charge.  Note, that  for the different  charges of the particles and bulk starting points $z_*$, particles probe different regions. In Fig.\ref{fig:world} we plot particles with different charges starting from the same point $z_*$.  The red curve corresponding to chargeless particle probes all regions and finally asymptotically falls on the horizon. The green,  blue and magenta  curves correspond to the charge value higher than critical.  The green one after probing the Rindler region for a finite time returns to the throat and oscillates, and the blue curve is confined to the outer region and never enter the throat. The magenta curve is confined to the throat and never enters the Rindler and outer regions. The universal chaotic behaviour corresponds to the particle confined to the Rindler region  with exponentially growing momentum at asymptotic future at  $t \rightarrow \infty$ (red curve).

\begin{figure}[h!]
\centering
\includegraphics[width=9cm]{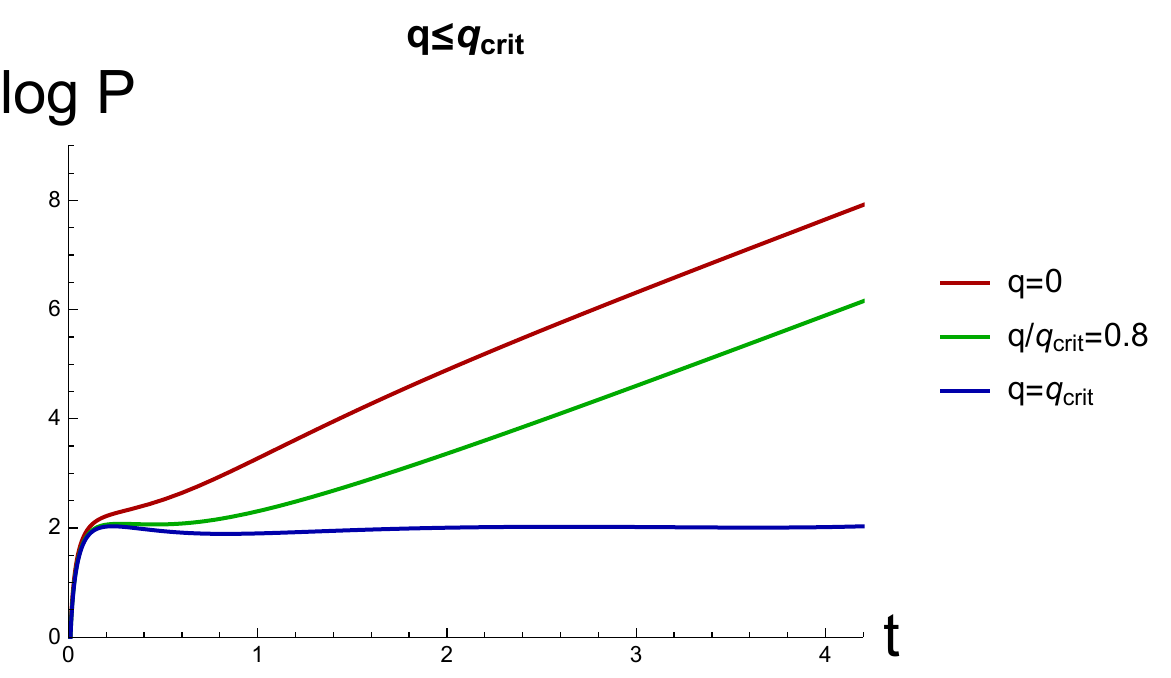}\,\,\,\,
 \caption{The momentum time dependence for different charge values and $\mu=1.3$, $d=3$, $z_h=1$, $z_*=0.1$. The red line corresponds to the neutral particle, the green one to $q/q_{crit}=0.8$ and the blue one to $q=q_{crit}$.  }
 \label{fig:nonextr-mom}
\end{figure}

\begin{figure}[h!]
\centering
\includegraphics[width=9cm]{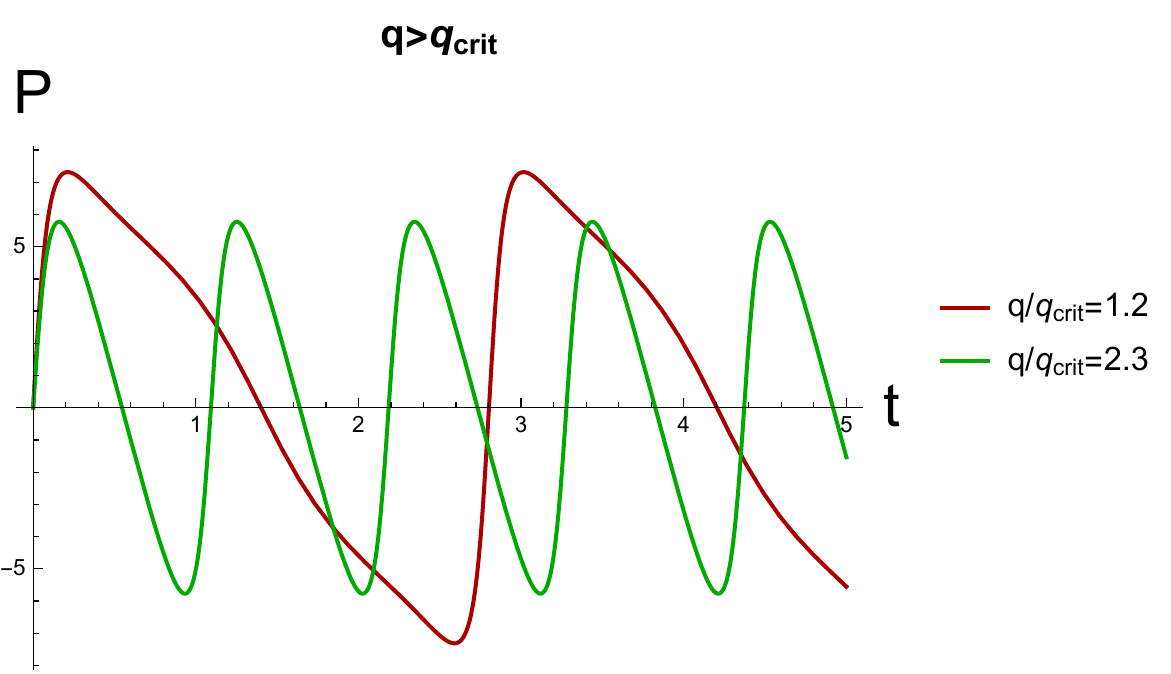}
 \caption{The momentum time dependence for different charge values and $\mu=1.3$, $d=3$, $z_h=1$, $z_*=0.1$. The red curve corresponds to $q/q_{crit}=1.2$ and the green one to $q/q_{crit}=2.3$.  }
 \label{fig:nonextr-mom-1}
\end{figure}

The radial momentum of the charged particle corresponding to  the metric \eqref{metr-1} is given by 
\bea 
&&p_z=\frac{1}{z f(z)}\frac{\dot z}{\sqrt{f(z)-\dot z^2/f(z)}}.
\eea 
In Fig.\ref{fig:nonextr-mom} and Fig.\ref{fig:nonextr-mom-1} we plot the time evolution of  the particle momentum   for  the charge  above and below the critical value. 

Note the following details concerning the momentum and, consequently, operator size growth following from the Fig.\ref{fig:nonextr-mom-1} and  Fig.\ref{fig:nonextr-mom}:
\begin{itemize}
\item The late time growth  of the particle momentum for all $q<q_{crit}$ is independent on $q$. The onset of this regime occurs at a later times for a smaller values of $1-q/q_{crit}$. This implies the scrambling time growth. This effect  is present for example in the random circuit models (see discussion for more details). 
\item For the critical charge the operator size stops evolving after some time. 
\item Above the critical value of the charge, the initial period of rapid linear growth is replaced by slow nonlinear growth, and the particle reaches its maximum velocity. Then a period of nonlinear deceleration occurs, which goes into fast linear deceleration and after a short nonlinear deceleration the particle reaches its minimum velocity. After slow growth, the particle reaches again a rapid linear growth, and so on. This oscillating behavour can be seen as the localization of the operator on some RG scale.
\item Note, that for the values of charge above the critical the momentum value could be negative.  It is natural to assume that the operator size corresponds to the absolute value of the momentum.

\end{itemize}

For the positive charge the repulsion between the black hole and particle is absent. The late time exponential asymptotic for the operator size evolution is the same as for the case of the neutral particle. However similarly to the \cite{Susskind2} the size value is large in comparison to the neutral particle case.   Note that all of these results hold for all  values of chemical potential below the extremal one. When the chemical potential is equal the extremal one the Rindler region is absent and the exponential growth of the momenta is absent too.

\section{The local energy scale}

In the revised version of \cite{Susskind2} authors proposed the modification of the momentum/size duality. The main motivation of the modification was to implement the renormalization group (RG) scale into this correspondence. The proposed modification is

\bea 
\text{\bf Operator size} &\longleftrightarrow &\beta \cdot {\bf p_z(t)},\label{II}\\
 \beta &\sim& \frac{1}{\partial_z f(z)},\nn
\eea
where it is assumed that the evolution of the operator takes place in the theory dual to the black hole with the blackening function $f(z)$. Then $\beta$ is associated to the thermal length on the energy scale $z(t)$. In another words the RG scale dependence is taken into account by the temperature of the black hole with the horizon placed at the particle bulk coordinate.

Now consider how the evolution of the size is changed after taking into account the RG scale dependence in comparison to size growth presented in Fig.\ref{fig:nonextr-mom} and Fig.\ref{fig:nonextr-mom-1}.

\begin{figure}[h!]
\centering
\includegraphics[width=6cm]{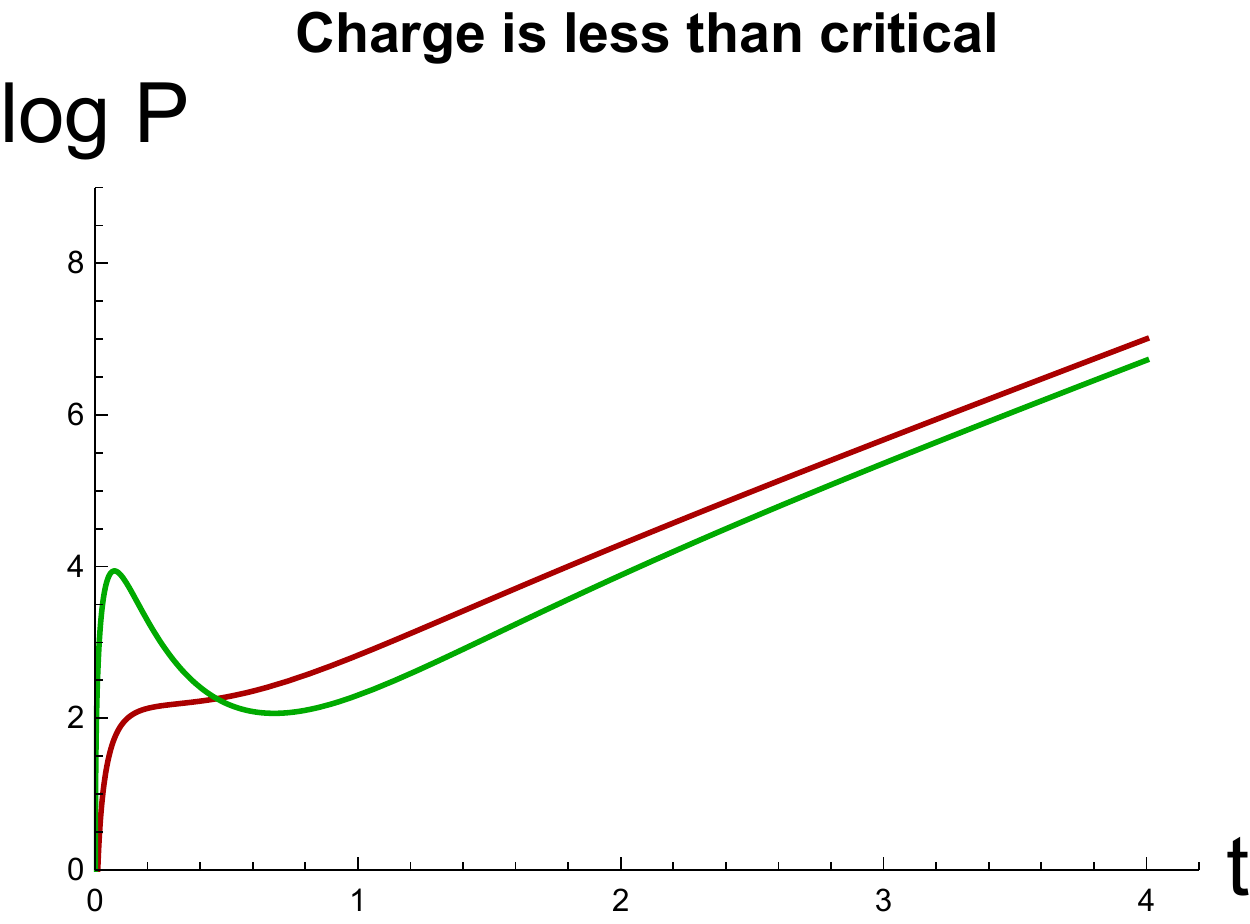}\,\,\,\,
\includegraphics[width=7.8cm]{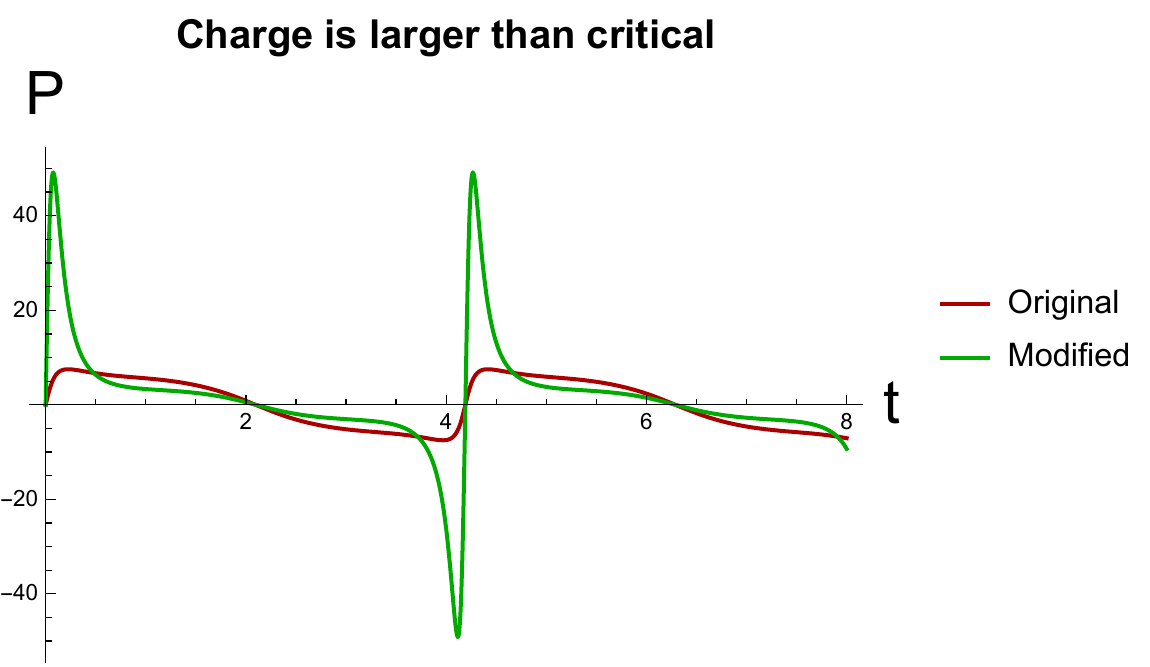}
 \caption{The momentum time dependence for different charge values and $\mu=1.3$, $d=3$, $z_h=1$, $z_*=0.1$. The left plot corresponds to the charge value $q=4$ (less than the critical one). The right plot corresponds to the charge $q=9$ (larger than the  critical one).  Red curves on both plots correspond to the duality without the inclusion of the energy scale. The green curves correspond to the modified duality.}
 \label{fig:mod}
\end{figure}

We see that all of the qualitative features of the size evolution are not changed. However the left plot in Fig.\ref{fig:mod} shows, that the inclusion of the energy scale to correspondence makes the size time dependence non-monotonous (green curve).

\section{Discussion and connection with SYK-like models} In this paper  we have  obtained that  the chemical potential lead to the operator size growth suppression for the charged operators. Also we discuss the RG scale modification of the momentum/size correspondence from \cite{Susskind2} and show that it does not change our main conclusions and results.
Our results rely on  the holographic proposal of  \cite{Susskind1}.
Now let us briefly list a few results observed in melonic theories at a finite chemical potential that support this observation.

 In \cite{Bhattacharya:2017vaz} it was shown, that in  the SYK model  with the complex fermions  introduction of the chemical potential  lead to the effect of chaos damping. There is a little difference with our results. In our case the full suppression occurs only at the infinite $\mu$ in contrast to the full suppression in the complex fermions SYK model appeared at finite value of $\mu$. This may be related with the large $q$ approximation used in \cite{Bhattacharya:2017vaz}.  

It is known \cite{Azeyanagi} that  a special large $D$ limit of  matrix quantum mechanics is dominated by the melonic diagrams. This fermionic model with a mass term, playing the role of the chemical potential, exhibits sharp phase transition that may be related to a chaos damping.

Another interesting model exhibiting some of these features is the random circuit model studied in \cite{Rakovszky:2017qit,Khemani:2017nda}. These works show, that the operator scrambling time becomes exponentially large in chemical potential if we consider the correlators with some conservation law. Also there could be some relation between the localization in random circuits constrained by $U(1)$ conservation considered in \cite{Pai:2018gyd} on one side and localization of operator size described by oscillating bulk particle.  

All these are in the accordance with the holographic example of $GR=QM$ discussed in this paper.

To conclude,  let us briefly summarize the results of the paper and future directions to investigate. In this paper we have considered the holographic model of the operator growth at a finite chemical potential. This model consists of the charged particle moving in the background of the charged black hole with the same charge sign. This holographic model shows that the chemical potential leads to the chaos suppression above the  critical charge. This is consistent with the chaos suppression observed in different melonic and random circuit theories mentioned above.

Also let us note a few future possible research directions.
The first one is to elaborate an analytic estimate for the charged operators and  quantitative tests with some quantum theory at finite temperature. For example, this includes a generalization of results from \cite{Size1}. Another direction is to explore holographic calculations of the size growth to  general backgrounds including hyperscaling-violating geometries
and the $dS$ case.

\section*{Acknowledgements}
Authors would like to thank Adam Brown, Matt Hodel, Nicholas Hunter-Johnes, Mikhail Katsnelson, Anatoly Polkovnikov  and Larus Thorlacius for discussions and correspondence. 
 The work is supported by the Russian Science Foundation (project 17-71-20154).


\begin{thebibliography}{} 




\bibitem{Susskind1} 
  L.~Susskind,
  ``Why do Things Fall?,''
  arXiv:1802.01198 [hep-th].


\bibitem{Susskind2} 
  A.~Brown, H.~Gharibyan, A.~Streicher, L.~Susskind, L.~Thorlacius and Y.~Zhao,
  ``Falling Toward Charged Black Holes,''
  arXiv:1804.04156 [hep-th].


\bibitem{Maldacena:2015waa} 
  J.~Maldacena, S.~H.~Shenker and D.~Stanford,
  ``A bound on chaos,''
  JHEP {\bf 1608}, 106 (2016)
    [arXiv:1503.01409 [hep-th]].

\bibitem{Shenker:2014cwa} 
  S.~H.~Shenker and D.~Stanford,
  ``Stringy effects in scrambling,''
  JHEP {\bf 1505}, 132 (2015)
    [arXiv:1412.6087 [hep-th]].

\bibitem{Size1} 
  D.~A.~Roberts, D.~Stanford and A.~Streicher,
  ``Operator growth in the SYK model,''
  arXiv:1802.02633 [hep-th].


\bibitem{Bhattacharya:2017vaz} 
  R.~Bhattacharya, S.~Chakrabarti, D.~P.~Jatkar and A.~Kundu,
  ``SYK Model, Chaos and Conserved Charge,''
  JHEP {\bf 1711}, 180 (2017)
   [arXiv:1709.07613 [hep-th]].
\bibitem{Azeyanagi} 
  T.~Azeyanagi, F.~Ferrari and F.~I.~Schaposnik Massolo,
  ``Phase Diagram of Planar Matrix Quantum Mechanics, Tensor, and Sachdev-Ye-Kitaev Models,''
  Phys.\ Rev.\ Lett.\  {\bf 120}, no. 6, 061602 (2018)
    [arXiv:1707.03431 [hep-th]].

\bibitem{Rakovszky:2017qit} 
  T.~Rakovszky, F.~Pollmann and C.~W.~von Keyserlingk,
  ``Diffusive hydrodynamics of out-of-time-ordered correlators with charge conservation,''
  arXiv:1710.09827 [cond-mat.stat-mech].
  
\bibitem{Khemani:2017nda} 
  V.~Khemani, A.~Vishwanath and D.~A.~Huse,
  ``Operator spreading and the emergence of dissipation in unitary dynamics with conservation laws,''
  arXiv:1710.09835 [cond-mat.stat-mech].

\bibitem{Pai:2018gyd} 
  S.~Pai, M.~Pretko and R.~M.~Nandkishore,
  ``Localization in fractonic random circuits,''
  arXiv:1807.09776 [cond-mat.stat-mech].

\end{thebibliography}
\end{document}